\documentclass[12pt]{article}
\usepackage{amsmath}
\usepackage{graphicx}
\usepackage{amssymb} 
\newcommand{\Mat}[1]{{{\boldsymbol{#1}}}}

\def\be{\begin{equation}}
\def\ee{\end{equation}}
\def\dd{\mathrm{d}}
\title{{\bf DIRAC EQUATION FROM THE HAMILTONIAN AND THE CASE WITH A GRAVITATIONAL FIELD}}

\bigskip
\author{
{\bf Mayeul Arminjon}\\
\small\it Dipartimento di Fisica, Universit\`a di Bari,\\
\small\it Via Amendola 173, I-70126 Bari, Italy.
\footnote{
\ On leave from Laboratoire ``Sols, Solides, Structures''
(Unit\'e Mixte de Recherche of the CNRS),
 BP 53, F-38041 Grenoble cedex 9, France.
} 
}
\date{ }

\begin{document}

\maketitle
\vspace{1cm}

\noindent Starting from an interpretation of the classical-quantum correspondence, we derive the Dirac equation by factorizing the algebraic relation satisfied by the classical Hamiltonian, before applying the correspondence. This derivation applies in the same form to a free particle, to one in an electromagnetic field, and to one subjected to geodesic motion in a static metric, and leads to the same, usual form of the Dirac equation---in special coordinates. To use the equation in the static-gravitational case, we need to rewrite it in more general coordinates. This can be done only if the usual, spinor transformation of the wave function is replaced by the 4-vector transformation. We show that the latter also makes the flat-space-time Dirac equation Lorentz-covariant, although the Dirac matrices are not invariant. Because the equation itself is left unchanged in the flat case, the 4-vector transformation does not alter the main physical consequences of that equation in that case. However, the equation derived in the static-gravitational case is not equivalent to the standard (Fock-Weyl) gravitational extension of the Dirac equation.\\



\section {INTRODUCTION}

Dirac's equation for a relativistic particle is unanimously recognized as a historical step, although it is now known \cite{Wightman1972} that several arguments promoting the Dirac equation against the then competing Klein-Gordon equation were not conclusive in view of the later developments. The main argument in Dirac's first paper on his equation \cite{Dirac1928a} was that according to which ``{\small the wave equation }[should be] {\small linear in $W$ }[the energy] {\small or $\partial /\partial t$, so that the wave function at any time determines the wave function at any later time.}" But any system of partial differential equations may be rewritten (in many ways) as a first-order system, {\it e.g.} by introducing partial derivatives of the unknowns as new unknowns. The Klein-Gordon equation may indeed be rewritten in the Schr\"odinger form vindicated by Dirac, by introducing a 2-component wave function \cite{FeshbachVillars1958}. Another argument noted that the time component of the Dirac current is positive, whereas that of the Klein-Gordon current is not and hence does not allow a probability interpretation. In the mean time, however, it has been found that the localization of quantum relativistic particles must appeal to different coordinate operators \cite{NewtonWigner1949}: therefore, ``in this respect the Klein-Gordon equation is neither better nor worse than the Dirac equation" \cite{Wightman1972}. It is now accepted that the Klein-Gordon equation describes spin-0 particles, whereas Dirac's applies to spin-1/2 particles.\\
\indent Since the Dirac equation remains extremely important today, the derivation of this equation is an important point. The textbooks ({\it e.g.} Bjorken and Drell \cite{BjorkenDrell1964}, Schulten \cite{Schulten1999}) follow closely Dirac's original derivation based on the factorization of the Klein-Gordon operator. There are also derivations ({\it e.g.} Ryder \cite{Ryder1988}) which assume notions such as spin and spinors, and which therefore, one might say, start from knowledge which Dirac's equation has the advantage of leading to naturally. On the other side, we find derivations which are based on an extraneous framework, quite remote from the problem solved by Dirac: Srinivasan and Sudarshan \cite{Srinivasan1996} introduce quaternion measurable processes, {\it i.e.}, families of quaternion measures; Close \cite{Close2002} shows that a Dirac-type equation describes torsion waves in an elastic solid; C\'el\'erier and Nottale \cite{CelerierNottale2002} start from scale relativity with its fractal space-time. In the present paper, the Dirac equation will be obtained more directly from the Hamiltonian of a classical relativistic particle. One advantage is that the same method, and indeed essentially the same equation, now applies to the cases of: a free particle; a particle in the electromagnetic field; and a particle in a static gravitational field. The gravitational Klein-Gordon equation has previously been obtained by this method \cite{B15}. Another point is that the method is based on an interpretation \cite{B15,A22} of the correspondence between a classical Hamiltonian and a quantum wave equation, which may be considered as a justification of that correspondence---whereas the latter is usually taken as an axiom. The fact that the derivation based on this interpretation/justification applies also to the gravitational case means that this approach to writing quantum wave equations in a gravitational field is more direct than the usual approach. The latter is inpired by Einstein's equivalence principle and aims at writing a covariant equation that coincides with the flat-space-time version if the coordinate system is such that, at the event considered, the metric is Minkowskian and the connection cancels. The application of this standard approach may be ambiguous, {\it e.g.} because covariant derivatives do not commute. Thus, the covariant generalization of the Klein-Gordon equation depends on an arbitrary parameter which multiplies the curvature scalar \cite{BirrellDavies}. In the case of the Dirac equation, the standard approach leads to the equation independently proposed by Fock \cite{Fock1929b} and Weyl \cite{Weyl1929b}, but this followed only after a rather involved analysis---as one may realize from the historical account \cite{Scholz2004}, as well as from the derivation of this extension of Dirac equation to curved space-time \cite{BadeJehle1953}.\\
\indent This complexity of the standard extension of the Dirac equation to the gravitational case is related to the special transformation law that is used for the Dirac wave function, namely, the spinor transformation. Especially in a curved space-time, spinor representations are related with elaborated differential-geometrical concepts. However, one may ask if this mathematical sophistication is both absolutely necessary, and physically adequate. In the present paper, it will be argued that the spinor transformation results from a contingent interpretation of the relativity principle in the context of the Dirac equation: another interpretation allows one to use the standard 4-vector transformation instead. Moreover, it will be found that, for the proposed version of the gravitational Dirac equation, only the 4-vector transformation can be used. The new interpretation of the relativity principle for Dirac's original (flat-space-time) equation does not change the physical consequences of the latter, at least not the direct ones. This is because these consequences, such as the emergence of spin and the precise prediction of the energy levels of hydrogen-type atoms, are those of the equation itself (see {\it e.g.} Schulten \cite{Schulten1999}), instead of being consequences of its transformation law under a Lorentz boost. However, the new derivation of the gravitational Dirac equation does lead to a different equation as compared with the standard version---this was already the case for the Klein-Gordon equation \cite{B15}. 

\section{ASSOCIATING A QUANTUM WAVE EQUATION WITH A CLASSICAL HAMILTONIAN}\label{Hamilton-wave-eqn}

Here, the results of a previous work \cite{A22} will be summarized and extended. A very important and rather mysterious feature of quantum mechanics (QM) is the classical-quantum correspondence, that associates a linear differential operator with a classical Hamiltonian, and that leads to regard energy and momentum as operators of this kind. The attempt \cite{A22} at interpreting the classical-quantum correspondence starts from remarks made by Whitham~\cite{Whitham} in the context of the theory of {\it classical waves}. Essentially, for any linear wave equation, one may define different ``wave modes," each of which is characterized by a {\it dispersion relation}, {\it i.e.} an explicit dependence of the frequency $\omega $ as a function of the spatial wave (co-)vector ${\bf k}$, $\omega =W({\bf k};X)$. (In the general case of heterogeneous propagation, the dispersion depends indeed on the space-time position $X$.) It turns out~\cite{Whitham} that, for a given wave mode, the wave vector ${\bf k}$ propagates along the bicharacteristics of a certain linear partial differential equation of the first order. When the latter equation is put into characteristic form, one obtains a {\it Hamiltonian system}, in which the Hamiltonian is none other than the dispersion relation $W$ defining the given wave mode of the wave equation considered. Now, with certain precautions which are made necessary by the existence of several wave modes, one may {\it recover the wave equation} from the dispersion relation alone. Hence one guesses that, under favourable conditions, one may associate a relevant wave equation with a classical Hamiltonian, as is done in QM.

\subsection{Dispersion equation and dispersion relations for a linear wave operator} \label{Dispersion}

In order to explicitly formulate this interpretation in a general-enough case, it is necessary to generalize somewhat Whitham's remarks \cite{A22}, especially in what regards the case of variable coefficients. The notion of a wave assumes that the wave function $\psi$ is defined over an open domain D in an extended configuration space $\mathrm{V}\equiv\mathsf{R}\times \mathrm{M}$, with $\mathrm{M}$ the $N$-dimensional configuration space. Consider a linear differential operator P defined ``on D" (in fact, P is defined on a suitable space of functions having D as common domain of definition). Let us assume a second-order operator for simplicity (this is not necessary, but it is enough for the application to QM):
\be \label{P-operator}
\mathrm{P}\psi \equiv a_0(X) +a_1^\mu (X)\partial _\mu \psi +a_2^{\mu \nu } (X)\partial _\mu \partial _\nu\psi,
\ee
where $X$, with coordinates $x^\mu (0 \leq \mu \leq  N)$, is the relevant point of the extended configuration space V. Since the operator is linear, it is relevant to consider solutions of the equation 
\be \label{wave-eqn}
\mathrm{P}\psi =0
\ee
in the form of sinusoidal waves:
\be \label{sinus-wave} 
\psi (X) = A(X)\, \mathrm{exp}[\, i\,\theta (X)],
\ee
where the phase $\theta (X)$ is a real function. For a function of the form (\ref{sinus-wave}) [whether it is a solution of (\ref{wave-eqn}) or not], we define the wave covector
\be \label{K}
{\bf K} = (K_\mu), \quad K_\mu \equiv \partial _\mu \theta.
\ee
Let a function of the form (\ref{sinus-wave}) have constant amplitude $A$ and be such that, at the point $X$ considered, we have
\be \label{elementary-wave}
\partial_\nu K_\mu (X)\equiv \partial_\nu \partial_\mu \theta (X) =0.
\ee
Substituting (\ref{sinus-wave}) into (\ref{P-operator}), and accounting for (\ref{elementary-wave}), one finds that {\it a such function obeys (\ref{wave-eqn}) at $X$ if and only if } 
\be \label{Ppsi=0,psi=elementary-wave}
\Pi_X({\bf K}).A=0,
\ee
where ${\bf K}\equiv {\bf K}(X)$, and where
\be \label{Pi}
\Pi_X({\bf K})\equiv a_0(X) +i\,a_1^\mu (X)K_\mu +i^2a_2^{\mu \nu } (X)K_\mu K _\nu.
\ee
If one considers equation (\ref{wave-eqn}) for scalar functions, $\psi (X) \in {\sf R}$ or $\psi (X) \in {\sf C}$, then the coefficients of P must be scalars, and (\ref{Ppsi=0,psi=elementary-wave}) for one non-zero $A$ is equivalent to $\Pi_X({\bf K})=0$. On the other hand, if $\psi (X)$ belongs to some finite-dimensional vector space E, say $\mathrm{E}={\sf C}^m$, then the coefficients of P are not necessarily scalars, but may also be matrices with $m$ rows and $m$ columns. In that case, $\Pi_X({\bf K})=0$ is equivalent to ``$\mathrm{P}.Ae^{ i\,\theta}=0 \ \mathrm{at} \ X$,  for every $A \in \mathrm{E}$" (assuming (\ref{elementary-wave})).
\footnote{
~A fixed system of coordinates has been assumed given on V: clearly, the coefficients of P are coordinate-dependent. If one allows for coordinate changes, one finds \cite{A22} that $\Pi_X$ is well-defined if and only if one can define a certain class of coordinate systems connected by ``infinitesimally linear" changes, that is,
\be \label{infinit-linear}
\frac{\partial ^2x'^\rho}{\partial x^\mu\partial x^\nu}=0
\ee
at the point $X(x_0^\mu)=X(x_0'^\rho)$ considered, and if one admits only those coordinate systems that belong to this class. In particular, if the space $\mathrm{V}$ is endowed with a pseudo-Riemannian metric $\Mat{\gamma}$, a relevant class is that of the locally geodesic coordinate systems (LGCS) at $X$, {\it i.e.}, $\gamma_{\mu \nu,\rho}(X)=0$ for all $\mu,\nu,\rho$. It is elementary to check that this is stable by a change (\ref{infinit-linear}). Conversely, if $(x^\mu)$ is an LGCS at $X$ for $\Mat{\gamma}$, and if one changes to coordinates $x'^\nu$, the Christoffel symbols of $\Mat{\gamma}$ at $X$ are, in the $x'^\nu$'s: $\Gamma^\mu _{\nu \rho }=\frac{\partial x'^\mu}{\partial x^\sigma }\frac{\partial ^2x^\sigma }{\partial x'^\nu \partial x'^\rho } $ \cite{Doubrovine1982}. If the new system is still an LGCS, they all cancel, hence the coordinate change must verify (\ref{infinit-linear}).
} 
The existence of {\it real} solutions ${\bf K}$ to the equation $\Pi_X({\bf K})=0$ is precisely the condition under which, at the point considered $X \in \mathrm{V}$, the linear equation (\ref{wave-eqn}) with (\ref{P-operator}) is indeed a {\it wave equation}. One checks easily that the correspondence between the linear operator P and the polynomial function with variable coefficients $\Pi$ is {\it one-to-one}, also in the case with matrix coefficients. The inverse correspondence, from $\Pi$ to P, consists simply of the substitution
\be \label{Pi-to-P}
K_\mu  \longrightarrow \partial _\mu/i. 
\ee
Suppose one is able to follow as function of $X$ the different roots of the {\it dispersion equation} $\Pi_X(\mathbf{K})=0$, seen as a polynomial equation for the frequency $\omega\equiv-K_0$, thus one is able to identify the different wave modes. Then it is possible to define different {\it dispersion relations} 
\be \label{dispersion}
\omega = W(K_1,...,K_N;X),
\ee
each of which gives the frequency $-K_0$ as a function of the spatial wave (co)vector $\mathbf{k}\equiv (K_1,...,K_N)$, {\it i.e.} the spatial part of $\mathbf{K}$, for the considered wave mode. For each of these functions $W$, an argument of Whitham (\cite{Whitham}, Sec. 11.5), reproduced in Ref. \cite{A22}, proves that, if a wave function (\ref{sinus-wave}) is such that the wave covector (\ref{K}) is a solution of Eq.~(\ref{dispersion}), then the modification of the spatial wave vector is governed by a Hamiltonian system with Hamiltonian $W$:
\footnote{ 
\ Whitham did not give a precise definition of the dispersion $W$ for equations having variable coefficients, although such are those for which the result (\ref{Hamilton-W-k})-(\ref{Hamilton-W-x}) is the most interesting.
} 
if the value of ${\bf k}$ is given at some point $X=(t,{\bf x})$, we have the coupled evolution given by\\
\be \label{Hamilton-W-k}
\frac{\dd k_j}{\dd t}= -\frac{\partial  W}{\partial x^j},
\ee
\be \label{Hamilton-W-x}
\frac{\dd x^j}{\dd t}= \frac{\partial  W}{\partial k_j} \qquad (j=1,...,N).
\ee

\subsection{The classical-quantum correspondence}

When they invented wave mechanics, de Broglie \cite{deBroglie1923b,deBroglie1925} and then Schr\"odinger \cite{Schroedinger1926a,Schroedinger1926b} assumed essentially that a Hamiltonian system describes in fact the ``skeleton" of a wave pattern associated with some linear wave equation, in exactly the same way as geometrical optics describes the trajectories of light rays, which are the skeleton of the underlying wave pattern. Geometrical optics corresponds to the ``nil wave length" limit for which, in the neighborhood of any point $X \in \mathrm{V}$, the wave (\ref{sinus-wave}) may be considered as a plane wave, that is
\be
A \simeq \mathrm{Constant\quad and}\quad \delta \theta \simeq  {\bf k.\delta x} -\omega \delta  t = K_\mu \delta x^\mu. 
\ee  
Thus, in this limit, Eq. (\ref{elementary-wave}) is satisfied and hence, by definition, substituting the sinusoidal wave function (\ref{sinus-wave}) into the wave equation (\ref{wave-eqn}) gives the dispersion equation $\Pi_X(\mathbf{K})=0$. Hence, in the geometrical optics limit, the wave function must obey one of the possible dispersion relations (\ref{dispersion}), and therefore the wave vector ${\bf k}$ must follow the Hamiltonian dynamics (\ref{Hamilton-W-k})-(\ref{Hamilton-W-x}). If, in this limit, the trajectories have to be governed by a classical Hamiltonian $H({\bf p},{\bf x},t)$, one is thus led to admit that the wave equation, which is searched for, must have as one of its dispersion relations a function $W$ whose Hamiltonian trajectories are just the solution trajectories of the classical Hamiltonian $H$, from which one starts. In order to ensure this, the most natural way \cite{A22} is to assume that the respective Lagrangians: $\Lambda$ for $W$ and $L$ for $H$, are proportional. Denoting $\hbar$ the proportionality constant, one gets simultaneously the relations of QM that relate energy with frequency and momentum with wave vector:
\be \label{E-omega}
H=\hbar W, \quad \mathrm{or} \quad E=\hbar \omega, 
\ee
\be \label{p-k}
\frac{\partial L}{\partial {\bf x}} =\hbar \frac{\partial \Lambda }{\partial {\bf x}}, \quad \mathrm{or} \quad {\bf p}=\hbar {\bf k}, 
\ee 
as well as the correspondence between a classical Hamiltonian and a wave operator. The latter correspondence is got by combining the correspondence (\ref{E-omega})-(\ref{p-k}) with (\ref{Pi-to-P}) ~\cite{A22}. Moreover, this interpretation provides a resolution to the ambiguity which is inherent in the correspondence, in the general case where the Hamiltonian contains terms that depend on both the (extended) ``position" $X=(t,{\bf x})$ and the canonical momentum ${\bf p}$. The rule to obtain the wave operator unambiguously is simple: just put the function of $X$ as a multiplying coefficient before the monomial in ${\bf p}$ –-- as the dispersion equation has to be a polynomial in ${\bf p}$ at fixed $X$~\cite{A22}. \\
\indent There remains a difficulty in the classical-quantum correspondence, however: the classical Hamiltonian $H({\bf p},{\bf x},t)$ is, in general, not a polynomial in ${\bf p}$ at fixed $X$ (just as the dispersion $W({\bf k};X)$ is, in general, not a polynomial in ${\bf k}$). An exception is the Hamiltonian of a particle subjected to a potential force field in classical mechanics,
\be \label{H-classical}
H({\bf p},{\bf x},t)= \frac{{\bf p}^2}{2m}+V(t,{\bf x}),
\ee
which leads thus directly to Schr\"odinger's equation. In a less particular case, it may still happen that the Hamiltonian is an {\it algebraic} function of the canonical momentum ${\bf p}$ (at fixed $X$): there exists a polynomial $Q(E,{\bf p};X)$, with complex coefficients, that cancels for $E=H({\bf p};X)$. For instance, for a relativistic particle,
\be \label{Q-SR-free}
Q(E,{\bf p})\equiv E^2-{\bf p}^2 c^2 -m^2 c^4=0 \quad \mathrm{if}\quad E=H({\bf p})
\ee  
in the free case, and
\be \label{Q-SR-em}
Q(E,{\bf p};X)\equiv (E-qV)^2-({\bf p}-\frac{q}{c}{\bf A})^2 c^2 -m^2 c^4=0 \quad \mathrm{if}\quad E=H({\bf p};X)
\ee
in the case of a charged particle (with charge $q$) in an electromagnetic potential $(V,{\bf A})$. We may, of course, choose for $Q$ the polynomial of the lowest possible order that cancels for $E=H({\bf p};X)$. [This remark concerns the general algebraic case: in the case of a relativistic particle, it is obvious that Eq.\ (\ref{Q-SR-free}) [or (\ref{Q-SR-em})] has the lowest possible order.] Then, in order to operate Schr\"odinger's wave mechanics so as to associate a linear operator with the classical Hamiltonian $H$, one may brutally apply to the algebraic equation $Q(E,{\bf p};X)=0$ the classical-quantum correspondence, {\it i.e.}, (\ref{E-omega})-(\ref{p-k}) followed by (\ref{Pi-to-P}). In the case of a relativistic particle, this leads immediately to the Klein-Gordon equation. It amounts to taking Eq.\ (\ref{Q-SR-free}) [or (\ref{Q-SR-em})] as the corresponding dispersion equation, up to the $\hbar $ factor and to the sign of $K_0=-\omega $, {\it i.e.} [in Cartesian space coordinates $(x^i)$]
\be \label{RelativisticDispersion}
\Pi({\bf K})\equiv K_0^2-c^2 K_i K_i -\frac{m^2c^4}{\hbar ^2}=0.
\ee
Let us select units such that $\hbar =c=1$ for simplicity, so that the correspondence (\ref{E-omega})-(\ref{p-k}) writes simply
\be \label{E-omega-p-k-hbar=1}
E=-K_0,\quad p_i=K_i,     
\ee
and the dispersion $W$ coincides with the Hamiltonian $H$. Then, let us define
\be \label{g^munu}
(g^{\mu \nu})\equiv (g_{\mu \nu})^{-1},  
\ee
with here (Minkowski metric in Galilean coordinates)
\be \label{g_munu-flat}
(g_{\mu \nu})\equiv  \mathrm{diag}(1,-1,-1,-1),
\ee
whence $(g^{\mu \nu})= \mathrm{diag}(1,-1,-1,-1)= (g_{\mu \nu})$ in the present case. Then we may rewrite (\ref{RelativisticDispersion}) in the following (generally-covariant) form:
\be \label{RelativisticDispersion-covariant}
\Pi({\bf K})\equiv g^{\mu \nu} K_\mu K_\nu -m^2=0.
\ee
\vspace{2mm}

We end this Section by noting that, if one admits that it is a wave structure which is the fundamental behaviour, then one may {\it a priori} expect that the classical-quantum correspondence is not always a sufficient tool to obtain a correct wave equation. However, in Subsect. \ref{Dispersion}, we associated one or several Hamiltonian systems to every linear wave equation. Conversely, if one starts from a Hamiltonian that is algebraic with respect to the canonical momentum ${\bf p}$, we just associated a wave equation to it, whereas it is likely that no linear wave equation can be naturally associated with a non-algebraic Hamiltonian. It seems thus that classical Hamiltonian systems are actually more general objects than are linear wave equations, and that, therefore, any relevant wave equation should be obtainable from a Hamiltonian by the classical-quantum correspondence. But, in the case of a relativistic particle, where an algebraic Hamiltonian is indeed there, {\it several} wave equations may in fact be associated with it. 

\section{THE DIRAC EQUATION} \label{Dirac}

In the ``algebraic" case described above, the algebraic equation [say Eq.\ (\ref{Q-SR-free})] will usually have other solutions than just the relevant Hamiltonian $H({\bf p};X)$. Correspondingly, the wave equation associated by the classical-quantum correspondence will have {\it too much solutions}, many of which shall be irrelevant to the dynamics described by the Hamiltonian $H$---thus, in the case of a relativistic particle, the Klein-Gordon equation is correct, but it is not enough to specify the behaviour, in other words it has too much solutions. One may then think of a {\it factorization} of the polynomial in the algebraic equation: if this factorization would indeed occur, one might hope that one of the factors is indeed relevant. Thus, in the case of Eq. (\ref{RelativisticDispersion-covariant}), we try to get:
\be \label{Pi-factorization}
\Pi({\bf K})\equiv g^{\mu \nu} K_\mu K_\nu -m^2= (a_0+i a_1^\mu K_\mu)(b_0+i b_1^\nu K_\nu).
\ee
[We put an $i$ factor before the first-order coefficients $a_1^\mu$ and $b_1^\nu$ for convenience: this is in order that $a_1^\mu$ and $b_1^\nu$ be directly the coefficients of the first-order derivatives in the associated operators, see Eq. (\ref{Pi-to-P}).] However, this factorization cannot occur with complex coefficients, since this would mean that $Q$ (or $\Pi $) is {\it not} the polynomial of the lowest possible order, contrary to our choice. [If $Q(E,{\bf p})=0$ for $E=H({\bf p})$, then, of course, one of the factors must cancel for $E=H({\bf p})$.] Hence, the unknown coefficients in the rightmost side of (\ref{Pi-factorization}) have to belong to some {\it larger} algebra A containing the complex field ${\sf C}$. Therefore, A does not need to be a commutative algebra, moreover the complex numbers shall be identified with the complex line ${\sf C}\,1_\mathrm{A}$ in A, since a larger algebra means, in particular, a vector space on ${\sf C}$ of dimension $d>1$. This means that (\ref{Pi-factorization}) should be more correctly rewritten as 
\be \label{Pi-factorization-Id-A}
(g^{\mu \nu} K_\mu K_\nu -m^2)\,1_\mathrm{A}= (a_0+i a_1^\mu K_\mu)(b_0+i b_1^\nu K_\nu).
\ee
Since we are interested only in the zeros of $\Pi ({\bf K})$, we may multiply the right-hand side (r.h.s.) by $b_0.a_0^{-1}$ on the left,
\footnote{\
In doing so, we impose the additional condition that $a_0$ must have an inverse, as will be indeed the case, see Eq. (\ref{a_0=-+im})---thus, we assume that $m\ne 0$.
} 
and thus we may assume that 
\be \label{a0=b0)}
a_0 = b_0.
\ee
The product decomposition (\ref{Pi-factorization-Id-A}) is then equivalent to the following equations:
\be \label{quadratic-terms}
2g^{\mu \nu}\,1_\mathrm{A}=-(a_1^\mu b_1^\nu + a_1^\nu b_1^\mu), 
\ee
\be \label{first-order-terms}
a_1^\mu a_0 + a_0 b_1^\mu = 0,
\ee
\be \label{constant-term}
-m^2 1_\mathrm{A}= a_0^2.
\ee
From Eq. (\ref{constant-term}), we see that we may impose that $a_0$ is (identifiable with) a complex number, $a_0 \in {\sf C}\,1_\mathrm{A}$, and that it is then
\be \label{a_0=-+im}
a_0=\epsilon im 1_\mathrm{A}, \quad \epsilon =\pm 1.
\ee
With this, it follows from Eq. (\ref{first-order-terms}) that
\be \label{b_1=-a_1}
a_1^\mu = -b_1^\mu \equiv \gamma ^\mu.
\ee
The relation (\ref{quadratic-terms}) that must hold for the quadratic terms is then
\be \label{Clifford}
\gamma ^\mu \gamma ^\nu + \gamma ^\nu \gamma ^\mu = 2g^{\mu \nu}\,1_\mathrm{A}.
\ee
This is the well-known relation defining the modified Clifford algebra. (The genuine Clifford algebra applies to the objects $d^0\equiv \gamma ^0, d^j\equiv i\gamma ^j$.) Thus, we have rewritten the original polynomial $\Pi ({\bf K})$ as the following product:
\be \label{FactorRelativisticDispersion}
\Pi ({\bf K}) 1_\mathrm{A}= \Pi _1({\bf K})\,\Pi _2({\bf K}), 
\ee
with
\be \label{Factors1-2-RelativisticDispersion}
 \Pi _1({\bf K})\equiv \epsilon im 1_\mathrm{A}+i\gamma ^\mu K_\mu, \quad \Pi _2({\bf K})\equiv \epsilon im 1_\mathrm{A}-i\gamma ^\mu K_\mu.
\ee
The corresponding dispersion equation $\Pi ({\bf K}) =0$ is satisfied as soon as either $\Pi _1({\bf K})=0$ or $\Pi _2({\bf K})=0$. Either possibility is a dispersion equation in its own right, and thus, by the correspondence (\ref{Pi-to-P}), is uniquely associated with a linear wave equation. If $\epsilon =+1$, the first possibility leads to the Dirac equation:
\be \label{Dirac-free}
(i\gamma ^\mu \partial_\mu -m1_\mathrm{A})\psi =0,
\ee
while the second one leads to its well-known associate,
\be \label{Dirac-free-associate}
(i\gamma ^\mu \partial_\mu +m1_\mathrm{A})\psi =0.
\ee
The reverse occurs, of course, if one chooses $\epsilon =-1$.\\
\indent In the case with an electromagnetic (e.m.) field, the correspondence (\ref{E-omega-p-k-hbar=1}) means rewriting the algebraic equation (\ref{Q-SR-em}) as the dispersion equation
\be \label{RelativisticDispersion-em}
\Pi^\mathrm{em}_X({\bf K})\equiv (-K_0-qV)^2- (K_i-qA^i)(K_i-qA^i) -m^2=0
\ee
in Cartesian space coordinates. Let us define (for a given value of $X$)
\be \label{K'}
K'_0 \equiv K_0+qV(X), \qquad K'_i \equiv K_i-qA^i(X),
\ee
that is, in general coordinates, simply
\be \label{K'}
K'_\mu  \equiv K_\mu + qA_\mu(X) \qquad (A^0\equiv V, \quad A_\mu  \equiv g_{\mu \nu }A^\nu).
\ee
We have $\Pi^\mathrm{em}_X({\bf K})=\Pi({\bf K}')$. Therefore, the factorization of the polynomial (\ref{RelativisticDispersion-em}) in the $K_\mu $'s, as the product of two first-order polynomials, is obtained by substituting ${\bf K}'$ for ${\bf K}$ into the factorization (\ref{FactorRelativisticDispersion}) of the polynomial $\Pi({\bf K})$:
\begin{eqnarray} \label{FactorRelativisticDispersion-em}
\Pi^\mathrm{em}_X({\bf K})\,1_\mathrm{A} & = & (\epsilon im 1_\mathrm{A}+i\gamma ^\mu K'_\mu)(\epsilon im 1_\mathrm{A}-i\gamma ^\nu K'_\nu) \nonumber\\
& = & [\epsilon im 1_\mathrm{A}+i\gamma ^\mu (K_\mu + qA_\mu)][\epsilon im 1_\mathrm{A}-i\gamma ^\nu (K_\nu+ qA_\nu)].
\end{eqnarray}
We have again factorized the dispersion equation---this time, that valid with an e.m. field. As for the free case, we may apply then the biunivocal correspondence (\ref{Pi-to-P}) to either of the two polynomials on the r.h.s. of (\ref{FactorRelativisticDispersion-em}). With the first polynomial, and with $\epsilon =+1$, we get
\be \label{Dirac-em}
[i\gamma ^\mu (\partial_\mu +iqA_\mu) -m1_\mathrm{A}]\psi =0,
\ee
which is the Dirac equation in an e.m. field.\\
\indent Thus, we derived the original Dirac equation from a factorization of that polynomial which provides an algebraic equation for the classical Hamiltonian, instead of factorizing the Klein-Gordon operator. One might say that this is a variant of the original and usual derivation, rather than a totally new derivation. This variant insists, more than does the usual derivation, on the algebraic aspects of the Dirac equation. Note, for example, that the product decomposition assumed {\it a priori} (\ref{Pi-factorization-Id-A}) is a completely general product of two first-order polynomials, whereas the two operators assumed in the usual factorization of the Klein-Gordon equation ({\it e.g.} Bjorken-Drell \cite{BjorkenDrell1964}, Schulten \cite{Schulten1999}) are quite particular. Moreover, the present derivation is based on our interpretation of the classical-quantum correspondence (Section \ref{Hamilton-wave-eqn}). Thus, the basic reason for which the Klein-Gordon equation is not the last word is that the algebraic equation (\ref{Q-SR-free}) [or (\ref{Q-SR-em})] obeyed by the Hamiltonian is second-order, hence it is tempting to try a factorization, with the hope of getting a more fundamental equation. Also, in this context, it is clear from the beginning that the correspondence (\ref{E-omega})-(\ref{p-k}) applies to the Hamiltonian energy $E=H({\bf p},{\bf x},t)$ with the corresponding canonical momentum ${\bf p}$, so that the derivation for the free case extends to that with an e.m. field, in a fully justified way. Another benefit will appear in Section \ref{Gravitational-case}.

\section{TRANSFORMATION OF THE DIRAC EQUATION} \label{Transformation}

Let $x'^\mu=F^\mu((x^\nu))$, or in short $X'=F(X)$, be any {\it admissible} tranformation of the space-time coordinates, the admissibility being a flexible notion which will be subjected to changes. We investigate the transformation of the wave function $\psi $, solution of the Dirac equation (\ref{Dirac-free}) or (\ref{Dirac-em}), that corresponds to $F$. In this Section, we shall consider as admissible the {\it linear} transformations, and merely the ones which belong to some subgroup G of the group ${\sf GL}(4,{\sf R})$ of all possible linear transformations. Thus, in the standard study of the transformation of the Dirac equation, one considers $\mathrm{G}={\sf O}(1,3)$, the Lorentz group. We ask that, after any tranformation $L\in$ G, the value, in the new coordinates, of the wave function at a given event depend linearly on its value $\psi (X)$ in the old coordinates, because the Dirac equation is linear. Further, we ask that the corresponding linear operator $S$ (which acts on the set of the values taken by the wave function) be the same for any event $X$, because here we are considering special relativity with its homogeneous space-time, this homogeneity being preserved by a linear transformation. Thus we ask that  
\be \label{psi'}
\psi '(X')={\sf S}(L).\psi (X),
\ee
for some operator function of $L$, $S={\sf S}(L)$. In fact, we know that the simplest realization of the algebra (\ref{Clifford}) is provided by $4\times 4$ complex matrices $\gamma ^\mu$, thus $\mathrm{A}= {\sf M}(4,{\sf C})$, so that the values $\psi(X) $ of the wave function are elements of ${\sf C}^4$. The regular operators acting on the vector space ${\sf C}^4$ form the group $\mathrm{H} = {\sf GL}(4,{\sf C})$, identifiable with that of the inversible $4\times 4$ complex matrices; but, of course, this does not mean that the matrices $S={\sf S}(L)$ cannot be restricted to some smaller group. From (\ref{psi'}), it follows immediately (by considering the inverse transformation $L'=L^{-1}$ or the composition $L=L_2.L_1$) that we must have
\be \label{S=representation}
\forall L \in \mathrm{G}, \quad {\sf S}(L^{-1})=[{\sf S}(L)]^{-1}, \qquad {\sf S}(L_2.L_1)={\sf S}(L_2).{\sf S}(L_1),
\ee
hence ${\sf S}$ must be a representation of G into H. Using (\ref{psi'}) and (\ref{S=representation})$_1$ in the Dirac equation (\ref{Dirac-free}), and multiplying on the left by $S\equiv {\sf S}(L)$, we get
\be \label{transformDirac-preliminary}
(iS\gamma ^\mu  \partial _\mu S^{-1} -m) \psi' =0.
\ee
Since $S$ is a constant here, and since $\partial /\partial x^\mu =L^\nu_\mu \,\partial /\partial x'^\nu \equiv L^\nu_\mu \,\partial '_\nu $, it follows that
\be \label{transformDirac}
(iL^\nu_\mu S\gamma ^\mu S^{-1}\, \partial '_\nu -m) \psi' =0.
\ee
The same argument applies with an electromagnetic (e.m.) field (covector $A_\mu$), leading to 
\be \label{transformDirac-em}
[iL^\nu_\mu S\gamma ^\mu S^{-1}\, (\partial '_\nu+iqA'_\nu) -m] \psi' =0.
\ee
At this stage, the standard approach would state that the Dirac equation must be Lorentz-invariant, and that {\it hence} the new matrices appearing in (\ref{transformDirac}) and (\ref{transformDirac-em}),
\be \label{gamma'}
\gamma '^\nu \equiv L^\nu_\mu S\gamma ^\mu S^{-1}, \qquad S\equiv {\sf S}(L),
\ee
must be equal to the starting matrices $\gamma ^\nu$. This leads to the spinor representation ${\sf S}= {\sf S}_\mathrm{spinor}$, which is defined for $L \in {\sf O}(1,3)$ and cannot be extended to ${\sf GL}(4,{\sf R})$ \cite{Weyl1929b,Scholz2004,Doubrovine1982}. \\
\indent Yes: the relativity principle demands that the equations are Lorentz-{\it covariant}---and this, without introducing extraneous quantities such as the velocity with respect to a preferred reference frame. However, as is well-known, it is only for a {\it scalar} that this means {\it invariance}. For instance, consider the relativistic equation of motion of a charged test particle in an e.m. field,
\be \label{charged-particle}
m\frac{\dd U^\mu}{\dd s} = q F^\mu _\nu U^{\nu}.
\ee
This equation is manifestly Lorentz-covariant; if one substitutes the ``absolute derivative" for the usual one on the left, it even becomes generally-covariant. In it, the 4-velocity $U^\mu$ tranforms like a (four-)vector, and the e.m. field $F^\mu _\nu$ transforms like a mixed tensor. To deduce from this transformation behaviour some hints for the Dirac equation, we begin by remembering that the Dirac wave function has four components. Hence we may adopt the same condensed notation for Eq. (\ref{charged-particle}) as for the Dirac equation, thus $U \equiv (U^\mu)$, $F \equiv (F^\mu _\nu)$: 
\be \label{charged-particle-condensed}
m\frac{\dd U}{\dd s} = q FU.
\ee
This makes it clear that, in the relativistic equation of motion of a charged particle, there is one matrix object $F$, which plays the role of a coefficient, very much as do the four $\gamma ^\mu$ matrices in the Dirac equation (\ref{Dirac-free}). And this matrix $F$ is {\it not} invariant in the transformation of this equation. Namely, after any linear transformation $L \in{\sf GL}(4,{\sf R})$, thus $x'^\mu=L^\mu _\nu x^\nu$, it transforms like this:
\be \label{F'munu}
F '^\mu _\nu =  \frac{\partial x'^\mu}{\partial x^\rho }\, \frac{\partial x^\sigma }{\partial x'^\nu } F^\rho _\sigma,
\ee
that is,
\be \label{F'}
F'=L F L^{-1}.
\ee
[Recall that this includes the case of a Lorentz boost: $L \in {\sf O}(1,3)$.] Therefore, we are allowed to do the same for the Dirac equation, {\it i.e.}, we are allowed to investigate a {\it relativistic} transformation of that equation in which the $\gamma ^\mu$ matrices would not be Lorentz-invariant. Moreover, account must be made for the fact that the $\gamma ^\mu$ 's are not univoquely fixed by the algebra (\ref{Clifford}): it is well-known that any similarity transformation
\be \label{similarity}
\tilde{\psi }= S\psi, \qquad \tilde{\gamma} ^\mu =  S\gamma ^\mu S^{-1}, \qquad S \in {\sf GL}(4,{\sf C})
\ee
is allowed. Now Eq. (\ref{gamma'}) says precisely that, up to the transformation of the partial derivatives, the new matrices are deduced from the old ones by the similarity transformation corresponding to the very linear transformation which affects the wave function. Moreover, we note that the new matrices (\ref{gamma'}) verify 
\be \label{gamma'mu.gamma'nu}
\gamma '^\nu \gamma '^\rho = L^\nu_\mu L^\rho _\sigma S\gamma ^\mu \gamma^\sigma S^{-1}, 
\ee
whence, by (\ref{Clifford}), the anticommutation relations
\be \label{Clifford'}
\gamma'^\nu \gamma'^\rho + \gamma'^\rho \gamma'^\nu = 2g'^{\nu \rho } \,1_\mathrm{A},
\ee
with
\be \label{g'munu}
g'^{\nu \rho} \equiv L^\nu_\mu L^\rho _\sigma g^{\mu \sigma }.
\ee
The latter, of course, is precisely the expression of the (contravariant) metric tensor after the general linear coordinate change characterized by matrix $L \in {\sf GL}(4,{\sf R})$. In the case of flat space-time, and if one starts from a Galilean coordinate system, {\it i.e.} one in which the metric has the Minkowskian form (\ref{g_munu-flat}), that Minkowskian form is conserved if and only if $L \in {\sf O}(1,3)$. This is true with the matrices (\ref{gamma'}), independently of the chosen representation ${\sf S}$---provided ${\sf S}$ is defined for $L \in {\sf O}(1,3)$. \\
\indent Let us summarize. If one considers linear coordinate transformations $L$ belonging to some subgroup G of ${\sf GL}(4,{\sf R})$, and if {\it any} representation ${\sf S}$ of G into ${\sf GL}(4,{\sf C})$ is available, then the free Dirac equation (\ref{Dirac-free}), as well as that with an e.m. field, Eq. (\ref{Dirac-em}), are form-invariant after the transformation defined by Eqs. (\ref{psi'}) for the wave function and (\ref{gamma'}) for the matrices $\gamma ^\mu $, Eqs. (\ref{transformDirac}) and (\ref{transformDirac-em}). If  $\mathrm{G} \supset {\sf O}(1,3)$, the anticommutation relation (\ref{Clifford}) is invariant ({\it i.e.} with the components of the metric being invariant) after a Lorentz boost, Eqs. (\ref{Clifford'}) and (\ref{g'munu}). The standard choice has been to impose that the matrices $\gamma ^\mu $ themselves are Lorentz-invariant, which leads to the spinor representation ${\sf S}_\mathrm{spinor}$. This is a peculiar interpretation of the relativity principle, as: {\bf i}) the (archetypically relativistic) equation of motion for a charged particle does contain a Lorentz-{\it non}-invariant matrix $F$, Eqs. (\ref{charged-particle-condensed}) and (\ref{F'}), and {\bf ii}) there is no privileged set of matrices $\gamma ^\mu $ among the infinitely many sets satisfying the anticommutation relation (\ref{Clifford}).\\
\indent An equally valid interpretation of the relativity principle allows us to select the {\it simplest} available representation:
\be \label{S=Id}
{\sf S} = \mathrm{Identity}, \qquad {\sf S}(L)=L,
\ee
which is defined over the whole linear group, $\mathrm{G}={\sf GL}(4,{\sf R})$. Thus, from Eqs. (\ref{psi'}) and (\ref{gamma'}), we find that the wave function tranforms like an usual 4-vector:
\be \label{psi'-4-vector}
\psi '(X')=L.\psi (X), \qquad \psi '^\mu = L^\mu_\nu \psi ^\nu,
\ee
and the Dirac matrices transform in the following way:
\be \label{gamma'-4-vector}
\gamma '^\mu \equiv L^\mu_\nu L\gamma ^\nu L^{-1},
\ee
which resembles the transformation (\ref{F'}) of the e.m. field matrix, up to the fact that here there are four matrices, and, due to the transformation of the partial derivatives, they are ``mixed" by the transformation. The transformation (\ref{psi'-4-vector})--(\ref{gamma'-4-vector}) works as well for the Dirac equation in an e.m. field (\ref{Dirac-em}). It seems that little, if anything, is changed to the direct physical consequences of the Dirac equation: the point to be emphasized is that the {\it equation}, hence its {\it solutions}, are unchanged. Thus, the non-relativistic limit of the equation with an e.m. field, leading to the Pauli equation involving the spin term; the exact solutions in a Coulomb field, including the splitting of the wave function into two components, each of which again includes a spin term, and leading to the energy spectrum of hydrogen-type atoms and to the explicit stationary states---all of this can be taken {\it verbatim} from Schulten \cite{Schulten1999}. Similarly, the propagator theory is unchanged, for it is based on the Green's function of the Dirac equation, which we leave unchanged; therefore, we might also recopy the Bjorken-Drell \cite{BjorkenDrell1964} analysis. The Foldy-Wouthuysen transformation \cite{BjorkenDrell1964,FoldyWouthuysen1950} is in fact a change of the unknown (wave) function, thus it does not involve a coordinate change and can be taken as it is. It may be misleading that the wave functions which are solutions of the Dirac equation are often called ``spinors:" whether or not a function obeys the equation in a given coordinate system, is independent of the transformation behaviour on changing the coordinate system. The transformation behaviour is defined once one chooses the representation ${\sf S}$ in Eqs. (\ref{psi'}) and (\ref{gamma'}). It would have been clearer if one would have reserved the word ``spinor" to designate the particular {\it representation} ${\sf S}_\mathrm{spinor}$, which leaves the $\gamma ^\mu$ matrices invariant.

\section{THE CASE WITH A GRAVITATIONAL FIELD} \label{Gravitational-case}

In Sect. \ref{Hamilton-wave-eqn}, the correspondence between a Hamiltonian and a wave equation has been, to some extent, justified, in the framework of an interpretation of that correspondence. To {\it derive} the Dirac equation from this correspondence, in the situation with a gravitational field, we have to ask if there is a Hamiltonian for the motion of a test particle in a gravitational field. According to the foregoing interpretation, we need a Hamiltonian in the original sense of Hamilton's mechanics, thus one in which the time coordinate $t$ is independent of the position ${\bf x}$ in the configuration space, the latter being the three-dimensional (frame-dependent) space, not the space-time \cite{B15}. (The same requirement is introduced, on somewhat different grounds, by Tagirov \cite{Tagirov1999}.) This excludes the ``super-Hamiltonian" $H[(p^\mu),(x^\nu)]$ \cite{MTW}. In Ref. \cite{B15}, it was shown that there is indeed a Hamiltonian for the geodesic motion of a test particle in a {\it static} space-time metric $\Mat{g}=(g_{\mu \nu})$, and that it is given by
\footnote{\ 
Recall that a static metric is one for which \cite{L&L}
\be \label{static-metric}
g_{\mu \nu}=g_{\mu \nu}((x^j)) \qquad \mathrm{and}\qquad g_{0j}=0. 
\ee
Bertschinger \cite{Bertschinger1999} states a Hamiltonian for geodesic motion in the case of a general metric, which Hamiltonian does coincide with (\ref{HamiltonStatic}) in the static case. In Ref. \cite{Tagirov1999}, a Hamiltonian is written for the case that the metric is given in normal Gaussian coordinates. 
} 
\be \label{HamiltonStatic}
H({\bf p},{\bf x})= [g_{00}(h^{jk}p_j p_k +m^2)]^{1/2},
\ee
the canonical momentum ${\bf p}$ being in fact the usual momentum, {\it i.e.}
\be
p_j = m\gamma _v h_{jk} v^k,
\ee
where $v^j \equiv (g_{00})^{-1/2}\dd x^j/\dd t$ is the velocity in the static reference frame, measured with local clocks, and where $\Mat{h}=(h_{jk})$ is the spatial metric in that frame and $(h^{jk})\equiv (h_{jk})^{-1}$, $\gamma _v\equiv (1-v^2)^{-1/2}$ being the Lorentz factor (with $v^2\equiv h_{jk} v^j v^k$). The time $t$ is the static time coordinate, which is unique up to a constant factor \cite{A16}. Thus, we have a Hamiltonian satisfying the algebraic relation
\be
E^2-g_{00}(h^{jk}p_j p_k +m^2)=0 \qquad \mathrm{for}\ E=H({\bf p},{\bf x}),
\ee
so that we may apply the same method as in Sect. \ref{Dirac}. We multiply the latter relation by $g^{00}=g_{00}^{-1}$, accounting for the fact that $h^{jk}=g^{jk} \ (j,k=1,2,3)$ and $g^{0 j}=0$. Using then the correspondence (\ref{E-omega-p-k-hbar=1}), we get the dispersion equation
\be \label{dispersion-static-gravitation}
g^{\mu \nu} K_\mu K_\nu -m^2 =0.
\ee
{\it This is identical to the dispersion equation (\ref{RelativisticDispersion-covariant}) valid for a free particle in flat space-time. Therefore, just the same factorization (\ref{FactorRelativisticDispersion})-(\ref{Factors1-2-RelativisticDispersion}) can be used as it is, leading to the same Dirac equation (\ref{Dirac-free}).} One difference is that, in the case with a static gravitational field, the metric cannot be reduced to the Minkowskian form (\ref{g_munu-flat}) by a coordinate change {\it preserving the static character of the metric}, that is \cite{A16}
\be \label{static-compatible-change}
x'^0 =ax^0, \qquad x'^j=\phi ^j((x^k)).
\ee
But it is only in ``static-compatible" coordinates [{\it i.e.}, ones in which the metric has the form (\ref{static-metric})] that the Hamiltonian (\ref{HamiltonStatic}) does rule the motion. This means that, {\it now, the Dirac equation (\ref{Dirac-free}) is with ``deformed" $\gamma^\mu$ matrices, satisfying the anticommutation relation (\ref{Clifford}) with a non-Minkowskian metric tensor $g^{\mu \nu}$.}\\
\indent At this point, we have to ask in which coordinate systems it is allowed to write that Dirac equation. In the flat-space-time case, we have shown that the trivial representation (\ref{S=Id}) may be used in the transformation [(\ref{psi'}),(\ref{gamma'})], which means that the wave function $\psi $ transforms like a 4-vector, Eq. (\ref{psi'-4-vector}). Since the latter transformation is defined for any coordinate change, we may now extend (\ref{psi'}) to {\it any} regular transformation $F$ of the space-time coordinates:
\be \label{psi'-general-F}
\psi '(X')=L(X).\psi (X), \qquad L(X)\equiv \nabla F(X) \quad (L^\mu _\nu \equiv \partial x'^\mu/\partial x^\nu).
\ee
Hence, we may study the transformation of the Dirac equation for a such general $F$, by redoing the reasoning that led to the transformed Dirac equation (\ref{transformDirac}), with here $S\equiv L(X)$. The only problematic step is that leading from (\ref{transformDirac-preliminary}) to (\ref {transformDirac}), precisely because here $S=L$ depends on $X$. This step cannot be made as is, unless we assume that the transformation $F$ satisfies Eq. (\ref{infinit-linear}) at the event $X(x_0^\mu)=X(x_0'^\rho)$ considered. If this condition is satisfied, we recover the Dirac equation (\ref{Dirac-free}) in the new coordinates, though the $\gamma ^\mu$ matrices have changed to (\ref{gamma'-4-vector}). Thus, we find again, in the case of the Dirac equation, the general limitation which is inherent in our interpretation of the correspondence Hamiltonian -- wave equation: namely, that one must consider coordinate changes satisfying condition (\ref{infinit-linear})---see Note 1 here, and see Ref. \cite{A22} (or \cite{B15}), $\S 2.1$, for details. This limitation on the coordinate system to apply the correspondence is physical: consider the covariant form (\ref{RelativisticDispersion-covariant}) of the dispersion equation for a relativistic particle, say in a flat space-time. In any coordinate system, this is unambiguously associated with just one linear wave equation, by the correspondence (\ref{Pi-to-P}):
\be \label{K-G-wrong}
(g^{\mu \nu} \partial _\mu \partial _\nu +m^2)\psi =0.
\ee
If we started from Galilean coordinates, with $(g^{\mu \nu})= \mathrm{diag}(1,-1,-1,-1)$, this is indeed the Klein-Gordon equation; but otherwise, {\it e.g.} with spherical space coordinates, this is a physically meaningless equation. Coming back to static gravitation, we have, therefore, to define a privileged class of static-compatible systems, exchanging by transformations that verify (\ref{infinit-linear}). There is only one natural such class in the case of a general static metric: the systems for which the time coordinate $x^0$ is the static time $t$ (defined up to a scale change), and which, moreover, are {\it locally geodesic}, at the spatial position ${\bf x}$ considered, for the {\it spatial} metric $\Mat{h}$ in the static reference frame, that is,
\be \label{class}
x^0 = at, \qquad h_{jk,l}({\bf x})=0 \quad (j,k,l\in \{1,2,3\}).
\ee
Thus, we must impose that the gravitational Dirac equation has the form (\ref{Dirac-free}) only in the coordinate systems satisfying conditions (\ref{class}). But it is a characteristic feature of curved space that condition (\ref{class})$_2$ cannot be verified in an open domain. Hence, we must find the form taken by the equation in more general systems.

\section{GENERAL TRANSFORMATION OF THE DIRAC EQUATION WITH 4-VECTOR WAVE FUNCTION} \label{General-transform-Dirac}

We transform the Dirac equation (\ref{Dirac-free}), assumed valid (possibly with deformed $\gamma ^\mu$ matrices) in some coordinate system $(x^\mu)$, to a general coordinate system $(x'^\nu)$, using the {\it 4-vector transformation} (\ref{psi'-general-F}) of the wave function. As already seen, this leads first to Eq. (\ref{transformDirac-preliminary}) with $S\equiv L(X)$. We get then, by differentiating $L^{-1}\psi '$ (and since $\partial _\mu =L^\nu_\mu \,\partial '_\nu $):
\be \label{transformDirac-general}
i\gamma'^\nu [\partial'_\nu \psi ' +L.(\partial'_\nu L^{-1}) \psi'] -m\psi '=0,
\ee
where the $\gamma '_\nu$'s are given by (\ref{gamma'-4-vector}). The new term writes more explicitly
\be \label{L^{-1}-term}
[L.(\partial'_\nu L^{-1}) \psi']^\mu =L^\mu _\rho \{\partial'_\nu [(L^{-1})^\rho _\sigma ]\}\psi '^\sigma =\frac{\partial x'^\mu}{\partial x^\rho }\,\frac{\partial ^2x^\rho }{\partial x'^\nu\partial x'^\sigma }\psi '^\sigma.
\ee
Using the transformation law that expresses the new Christoffel symbols $\Gamma '^\mu_{\nu \sigma }$ as function of the old ones $\Gamma ^\rho _{\lambda \theta }$ \{{\it e.g.} Ref. \cite{Doubrovine1982}, Sect. 28, Eq. (25)\}, this is
\be \label{L^{-1}-term-final}
[L.(\partial'_\nu L^{-1}) \psi' ]^\mu =\left( \Gamma '^\mu_{\nu \sigma } - \frac{\partial x'^\mu}{\partial x^\rho }\,\frac{\partial x^\lambda }{\partial x'^\nu}\frac{\partial x^\theta }{\partial x'^\sigma }\Gamma ^\rho _{\lambda \theta }\right)\psi '^\sigma.
\ee
Thus, if one assumes that (\ref{Dirac-free}) is valid in a ``freely falling" system, such that $g_{\mu \nu,\rho} =0$ for all $\mu ,\nu $ and $\rho $, and hence with all $\Gamma ^\rho _{\lambda \theta }$ zero, he obtains from (\ref{transformDirac-general}) and (\ref{L^{-1}-term-final}) the following form of the equation in the general system $(x'^\nu)$:
\be \label{Dirac-general-equivalence principle}
\left(i\gamma'^\nu D'_\nu -m\right)\psi '=0,\qquad (D'_\nu \psi ')^\mu \equiv \psi '^\mu_{;\nu} \equiv \partial '_{\nu}\psi'^\mu + \Gamma '^\mu_{\nu \sigma }\psi'^\sigma.
\ee
In words: the standard way of adapting the Dirac equation to gravitation, {\it i.e., via} the equivalence principle, is compatible with the 4-vector transformation of the wave function---it leads to write the equation with the {\it usual covariant derivative of 4-vectors,} and with $\gamma ^\mu$ matrices that change according to (\ref{gamma'-4-vector}) after a coordinate change.\\ 
\indent However, our specific assumption is that the Dirac equation (\ref{Dirac-free}) is valid, indeed with deformed $\gamma ^\mu$ matrices, {\it in a static-compatible system verifying (\ref{class})}---because, in any such system, we derived (\ref{Dirac-free}) from the classical Hamiltonian (\ref{HamiltonStatic}). Using Eqs. (\ref{transformDirac-general}) and (\ref{L^{-1}-term-final}), we may rewrite this in any new coordinate system $(x'^\nu)$ as
\be \label{transformDirac-general-Delta}
\left(i\gamma'^\nu \Delta '_\nu -m\right)\psi '=0,\quad (\Delta '_\nu \psi ')^\mu \equiv \psi'^\mu_{,\nu} + \left( \Gamma '^\mu_{\nu \sigma } - \frac{\partial x'^\mu}{\partial x^\rho }\,\frac{\partial x^\lambda }{\partial x'^\nu}\frac{\partial x^\theta }{\partial x'^\sigma }\Gamma ^\rho _{\lambda \theta }\right)\psi '^\sigma.
\ee
In the static system $(x^\mu)$ verifying (\ref{class}), the non-zero terms of the connection are ({\it e.g., cf.} Ref. \cite{A16}, $\S $3.2)
\be \label{static-LGCS-connection}
\Gamma ^0_{0j}=\Gamma ^0_{j0}=\frac{1}{2} \frac{g_{00,j}}{g_{00}}, \qquad \Gamma ^j_{00}=\frac{1}{2}h^{jk}g_{00,k}.
\ee
The second term inside the bracket on the r.h.s. of Eq. (\ref{transformDirac-general-Delta})$_2$ just transports this old connection to the new system $(x'^\nu)$, as a $\left(^1_2 \right)$ tensor: since the connection does not cancel in the system $(x^\mu)$, this transported connection cannot cancel as a whole, in any other system $(x'^\nu)$. If we take a freely-falling system as the new system $(x'^\nu)$, we have all $\Gamma '^\mu_{\nu \sigma} $ zero. Hence, the equation (\ref{transformDirac-general-Delta})$_1$ obtained in this system is not the usual Dirac equation (\ref{Dirac-free})---in contrast with the gravitational Dirac equation obtained from the equivalence principle, be it that based on the 4-vector transformation [Eq. (\ref{Dirac-general-equivalence principle})] or the standard equation \cite{Fock1929b,Weyl1929b}. Thus, {\it Eq. (\ref{transformDirac-general-Delta})$_1$ is not equivalent to the extension of the Dirac equation obtained from the equivalence principle, independently of the spinor or vector transformation of the wave function which is chosen for the latter.} Let us now consider as the new system $(x'^\nu)$ a {\it static-compatible} one, thus one deduced from $(x^\mu)$ by a purely spatial change (\ref{static-compatible-change})$_2$. [We may forget the trivial scale change (\ref{static-compatible-change})$_1$.] After the tensorial transport of the connection, defined above, the non-zero terms turn out to keep the same expression [the right-hand sides in (\ref{static-LGCS-connection})] in the system $(x'^\nu)$. We may then rewrite $(\Delta '_\nu \psi ')^\mu$ explicitly, without any reference to the old system (and hence omitting the primes):
\be \label{Delta_0^0}
(\Delta _0 \psi )^0 = \psi ^0_{;0}-\frac{1}{2} \frac{g_{00,k}}{g_{00}}\psi ^k,
\ee
\be \label{Delta_j^0}
(\Delta _j \psi )^0 = \psi ^0_{;j}-\frac{1}{2} \frac{g_{00,j}}{g_{00}}\psi ^0,
\ee
\be \label{Delta_0^j}
(\Delta _0 \psi )^j = \psi ^j_{;0}-\frac{1}{2}h^{jk}g_{00,k}\psi ^0,
\ee
\be \label{Delta_k^j}
(\Delta _k \psi )^j = \psi ^j_{;k},
\ee
where $\psi ^\mu_{;\nu}$ is the covariant derivative of a 4-vector, Eq. (\ref{Dirac-general-equivalence principle})$_2$. Thus, in any static-compatible coordinate system, we may write our version of the Dirac equation in a static gravitational field as
\be \label{Dirac-static-general}
\left(i\gamma^\nu \Delta _\nu -m\right)\psi =0,
\ee
where the components $(\Delta _\nu \psi )^\mu$ are given by Eqs. (\ref{Delta_0^0})-(\ref{Delta_k^j}). This is covariant by any static-compatible coordinate change (\ref{static-compatible-change}).

\section{CONCLUSION}

We use an interpretation of the classical-quantum correspondence \cite{A22} based on the mathematical relationship between a linear wave operator and its dispersion relation(s), and also based on the idea of de Broglie and Schr\"odinger, according to which classical Hamiltonian mechanics describes the skeleton of an underlying wave pattern. Integrating these two aspects provides a solution \cite{A22} to the ambiguity of this correspondence, in the case that the Hamiltonian contains mixed terms with position ${\bf x}$ and canonical momentum ${\bf p}$. That case is relevant to the situation with a gravitational field \cite{B15}.\\
\indent However, for a relativistic particle, the classical Hamiltonian $H$ is not a polynomial in ${\bf p}$ (at fixed ${\bf x}$ and $t$), but an algebraic function of it. The corresponding algebraic relation has another solution, which is $E'=-H$, and which does not describe the same dynamics as that obtained with $H$. This leads to the idea that, in factorizing that algebraic relation, one might get a more fundamental wave equation. The factorization does occur with coefficients in the larger algebra of the $4\times 4$ matrices, and leads to the Dirac equation. Of course, it turns out that it also has negative-energy solutions, but nevertheless it is indeed more fundamental in that it describes more important particles than does the Klein-Gordon equation. The main interest of the proposed approach is that the same derivation applies to three different cases: that of a free particle, that of a particle in an e.m. field, and that of a particle subjected to geodesic motion in a static gravitational field. In the latter case, there is indeed a Hamiltonian. Thus, our static-gravitational version of the Dirac equation is derived from the classical Hamiltonian after factorization of the dispersion equation [Eqs. (\ref{FactorRelativisticDispersion}) and (\ref{dispersion-static-gravitation})], just in the same way as we derive the flat-space-time Dirac equation. Hence, this gravitational Dirac equation follows as directly from wave mechanics as does the flat-space-time Dirac equation, whereas the standard (Fock-Weyl) version just extends the flat version using the equivalence principle. It also implies that all solutions of this proposed version are solutions of the static-gravitational Klein-Gordon equation associated \cite{B15} with the dispersion (\ref{dispersion-static-gravitation}). In contrast, it is not usually the case that a solution to the Fock-Weyl extension of the Dirac equation obey either of the generally-covariant extensions of the Klein-Gordon equation. This confirms that the proposed gravitational Dirac equation is not equivalent to the standard extension, as was proved after Eq. (\ref{static-LGCS-connection}).\\
\indent A surprising result of this work is that the standard 4-vector transformation may be applied to the wave functions $\psi $ obeying the Dirac equation---provided one accepts that the $\gamma ^\mu$ matrices are not invariant in the transformation of the equation. Oviously, an equation involving a matrix object may be covariant without the matrix being invariant: a relevant example is the relativistic equation of motion for a charged particle in an e.m. field, Eqs. (\ref{charged-particle-condensed}) and (\ref{F'}). Furthermore, the set of the $\gamma ^\mu$ matrices is not uniquely defined, since the anticommutation relation (\ref{Clifford}) admits an infinity of solution sets. In the case of curved space-time, a coordinate change does anyway change the $\gamma ^\mu$ matrices, since it changes the metric coefficients in that relation (\ref{Clifford}). Thus, the proposed 4-vector behaviour of $\psi $, associated with the transformation (\ref{gamma'-4-vector}) for the $\gamma ^\mu$'s, which does leave the Dirac equation form-invariant, is formally admissible.
\footnote{\ 
The possibility of keeping the 4-vector behaviour is alluded to in Ref. \cite{BadeJehle1953}, p. 715. A ``four-vector behavior of the Dirac bispinor" is vindicated by Bell {\it et al.} \cite{BellCullerneDiaz2000}, in the framework of quaternion Dirac equation. But Eq. (3) of Ref. \cite{BellCullerneDiaz2000} is not the standard 4-vector transformation [Eq. (\ref{psi'-4-vector}) here], and when the former applies, it applies to transform an equation which is not equivalent to Dirac's, in contrast with the present study. In the present work, the 4-vector transformation arose in the study of the gravitational case.
}
As to the observational aspect: we argue that, since the flat-space-time Dirac equation itself is the same, nothing is changed to its closest applications (the emergence of spin and the energy levels of hydrogen-type atoms), neither to those based on the Feynman propagator \cite{BjorkenDrell1964}, which is a Green's function of the Dirac equation (\ref{Dirac-free}) or (\ref{Dirac-em}).  Since the Dirac equation (especially after it is adapted for a quantum field) has applications in many parts of basic physics, however, it is hard to state in advance that no physical consequence is changed.\\
\indent In the static-gravitational case, the direct application of the classical-quantum correspondence, which has been done in Sect. \ref{Gravitational-case}, would not make sense with the usually-used (spinor) transformation. Indeed, the latter is restricted to the Lorentz group, hence it could not be used to rewrite the equation in any static-compatible system, by using a general spatial coordinate change, as it had to be done in Sect. \ref{General-transform-Dirac}. It remains to investigate the differences that could occur in the weak-field limits of the two gravitational extensions of the Dirac equation: the present one (limited to a static field), and the standard one \cite{Fock1929b,Weyl1929b}, for which investigations of the weak-field limit already exist (Refs. \cite{Obukhov2001,KieferWeber2005}, and references therein). Perhaps, such differences might be observable in the future, {\it e.g.} in experiments on ultra-cold neutrons, such as transmission measurements through a horizontal slit in the Earth's gravitational field \cite{Nesvizhevsky2002,Nesvizhevsky2003}.

\bigskip

\noindent {\bf Acknowledgement.}  I am grateful to F. Selleri, F. Romano, and to the late C. De Marzo, for their hospitality in Bari. Additional thanks are due to Franco Selleri for his helpful remarks on this work.

\end{document}